\documentclass[aps,twocolumn]{revtex4}

\input epsf
\usepackage{amsmath}
\usepackage{amsfonts}
\usepackage{amssymb}%

\newcommand{\vev}[1]{\langle #1 \rangle}

\newcommand{\GeV}{{\rm\ GeV}}
\newcommand{\TeV}{{\rm\ TeV}}

\newcommand{\checker}[1]{#1_v}

\newcommand{\ZZ}{{\bf Z}}

\begin{document}

\title{
Possible Effects of a Hidden Valley on
Supersymmetric Phenomenology
}

\author{Matthew J. Strassler
}
\affiliation{Department of Physics,
P.O Box 351560, University of Washington,
Seattle, WA 98195\\
}
\begin{abstract}{A hidden valley sector may have
a profound impact on the classic phenomenology of supersymmetry.  This
occurs if the LSP lies in the valley sector.  In addition to reducing
the standard missing energy signals and possibly providing displaced
vertices (phenomena familiar from gauge-mediated and
R-parity-violating models) it may lead to a variable multiplicity of
new neutral particles, whose decays produce soft jets and/or leptons,
and perhaps additional displaced vertices.  Combined, these issues
might obscure supersymmetric particle production from search
strategies used on current Tevatron data and planned for the LHC.
The same concerns arise more generally for any model that has a
symmetry (such as T-parity or KK-parity) realized nontrivially in both
the standard-model and the hidden-valley sectors.  Possible
strategies for experimental detection are discussed, and
the potential importance of the LHCb detector is noted.  }

\end{abstract}

\maketitle

As analysis of Tevatron data proceeds and preparations for the Large
Hadron Collider (LHC) continue, it is important to explore the
phenomenology of experimentally-challenging models, so as to assure no
phenomena are overlooked.  Recent work on late-decaying gluinos
\cite{split, D0split}, and on multi-bottom or multi-tau decays of a
Higgs \cite{NMSSM,WeirdHiggs,4tau} are in this spirit.  The present
article is aimed at continuing this effort.  A class of
``hidden-valley'' models, in which a wide variety of new neutral
resonances may arise, was presented in \cite{hidval}.  It was also
argued that looking for neutral long-lived particles should be a high
priority in Higgs searches \cite{higgs}.  Here it is shown how the
hidden valley models, which can be attached without much experimental
constraint to models that solve the hierarchy problem, including
supersymmetry, extra dimensions and the little Higgs, can drastically
change the phenomenology usually associated with those solutions.


The standard search strategy for supersymmetry (SUSY) --- if R-parity
is conserved --- involves its missing energy signal.  This is
associated with the presence, in each event, of two of the lightest
supersymmetric particle (LSP), a stable neutral particle.  However,
even with conserved R-parity, this signal may be reduced below the
expected level, if the lightest standard model (SM) superpartner,
which we will call the ``LSsP'', is not in fact stable.  In this case
the decay of the two LSsP's generates a ``tagging'' signal in
essentially every supersymmetric event.  This occurs in several
contexts.  In low-scale gauge mediated SUSY-breaking (GMSB) models,
the LSsP is heavier than the gravitino, which is actually the LSP.
The LSsP has a lifetime that is (from the point of
experimental detection) largely
unconstrained; typical decays to the gravitino plus visible
particles may occur promptly, anywhere within the volume of the detector,
or outside the detector.  Other examples of theories with similar
phenomena include models with degenerate Wino or Higgsino LSPs
\cite{degenerate}, models with light hidden-sector singlets
\cite{martin}, and R-parity violating (RPV) models \cite{RPV}.  In
the latter class of models, the MET signature may be completely
absent. Phenomenological signals of interest include
metastable neutralino LSsP's decaying to a photon, $Z$ boson or higgs
boson plus missing transverse energy (MET),
\cite{neutralinoLSsP,LSsPtoZh}; metastable tau sleptons producing a
tau plus MET, possibly generating a track with a kink \cite{staukink};
three-body selectron or smuon decays \cite{threebodyslep}; metastable
gluinos producing a jet or jets plus MET, possibly with a track and
possibly at a displaced vertex \cite{split}; track stubs from
degenerate gauginos \cite{degenerate}; or even a metastable top squark
\cite{stopLSsP} discoverable as a charged track or neutral particle
ending at a vertex with a jet and often a lepton.  Many interesting
search strategies have been proposed, studied, and in some cases
carried out \cite{studies,timing,cdfZsearch,mumu,D0split}.  These include
searches for non-pointing and/or late photons or Z's, large
negative-impact-parameter jets, out-of-time decays, etc.  Final
states, compared to the minimal-supergravity scenario, involve a
reduced missing energy signal plus two sets of LSsP decay products.
In all of these contexts, the decays of the LSsP reduce the missing
energy signal, provide a ``tag'' that in some cases can be used to
identify the supersymmetric events, generate new and often
diagnostically-useful kinematic distributions, and potentially produce
late decays that in some cases are easily seen and in other cases make
detection an experimental challenge.

Hidden valley models can very
easily produce LSsP decays that share many features with those listed
above --- a reduced missing energy signal and two sets of LSsP decay
products.  This is a general possibility whenever the LSP lies in the
valley sector.  Any production of SM superpartners leads to a cascade
down to two LSsPs. Then each LSsP will decay to the v-sector, and its
decay will therefore participate in (and thus both reveal and be
confused by) whatever dynamics occurs in that sector.  For this
reason, hidden valley models can show additional subtle and
experimentally relevant differences from other SUSY models.

\begin{figure}[htbp]
  \begin{center}
    \leavevmode
     \epsfxsize=.45 \textwidth
     \hskip 0in \epsfbox{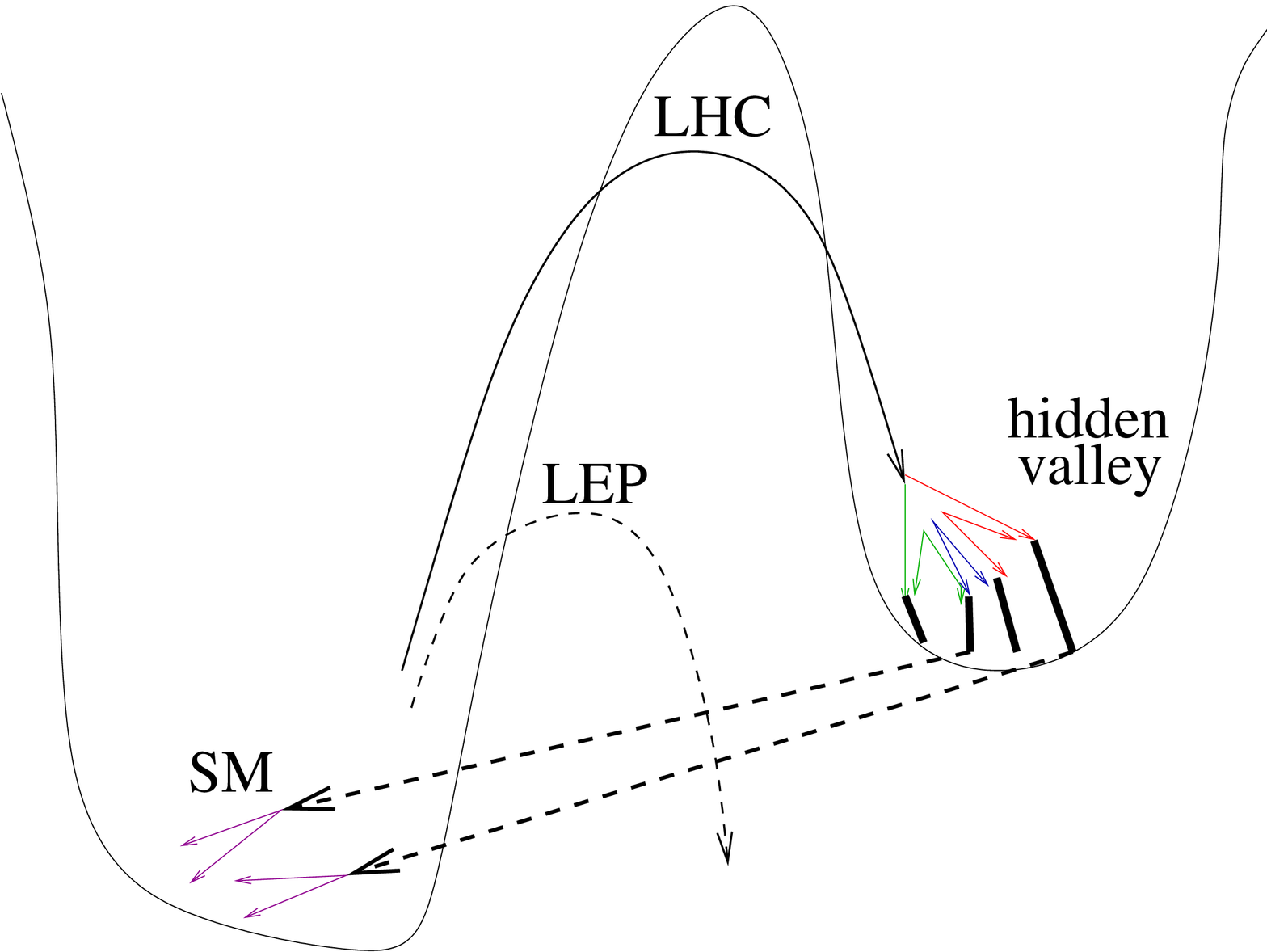}
  \end{center}
\caption{Schematic view of production and decay of v-hadrons.
While LEP was unable to penetrate the barrier separating the sectors,
LHC may easily produce v-particles.
These form v-hadrons, some of which decay to standard model
particles.}
\label{fig:hidval}
\end{figure}

\begin{figure}[htbp]
  \begin{center}
    \leavevmode
     \epsfxsize=.45 \textwidth
     \hskip 0in \epsfbox{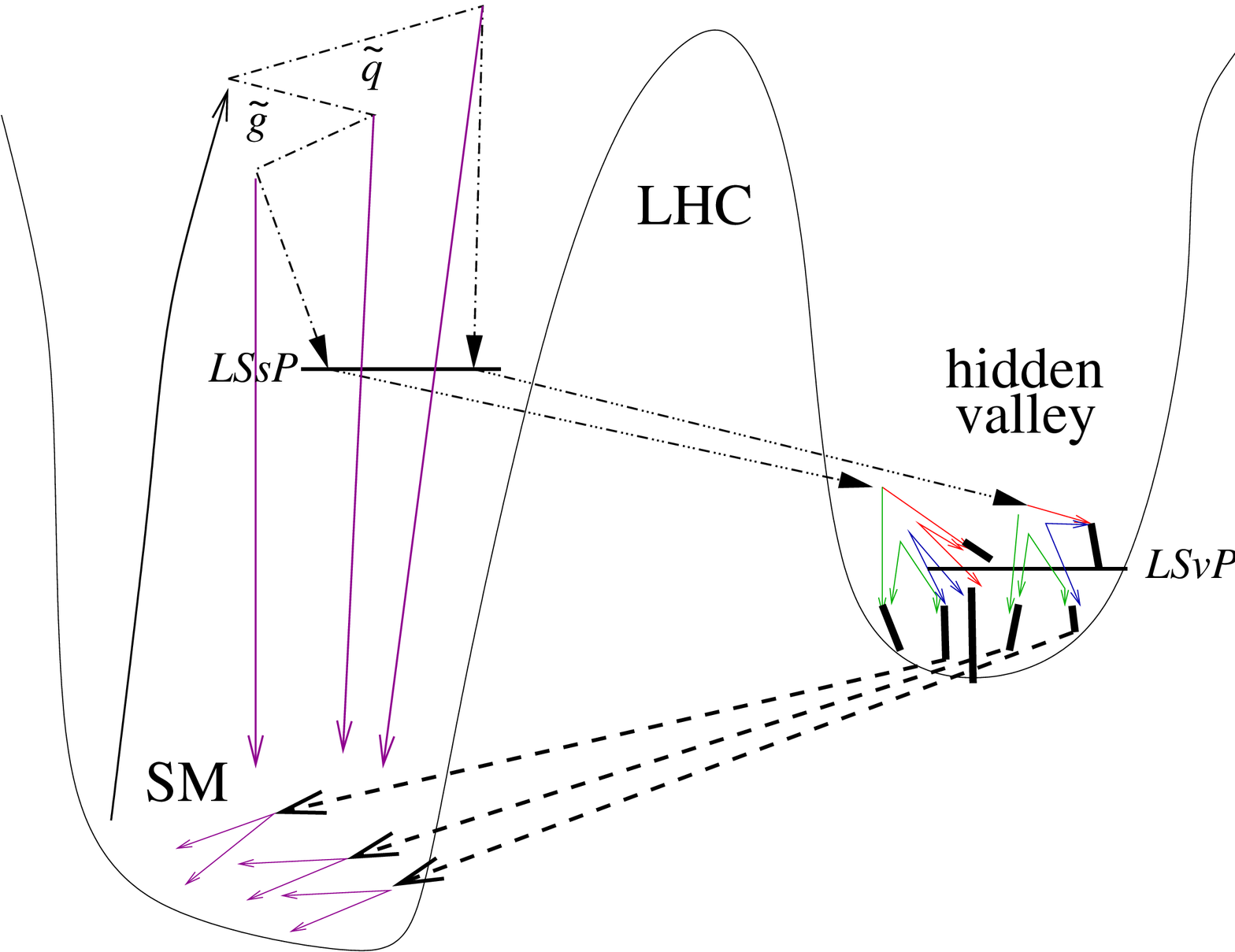}
  \end{center}
\caption{Schematic view of production and decay of SM superpartners.
Each superpartner decays to hard jets/leptons and an LSsP;
the LSsP then decays to an LSvP plus other v-hadrons, some
of which decay to softer jet/lepton pairs.
}
\label{fig:valleySUSY}
\end{figure}

In a typical hidden-valley model, the valley sector (``v-sector'') has
its own matter --- valley quarks, or ``v-quarks'', and their
superpartners, v-squarks --- and its own gauge group, with v-gluons
and v-gluinos.  In confining hidden valley models, these v-particles
are confined and form v-hadrons.  In \cite{hidval}, the possibility
was considered that v-particles are produced via a $Z'$ decay; some of
the v-hadrons produced in v-hadronization can then decay back to
standard model particles, via an intermediate state $Z'$ or Higgs
boson.  This is illustrated schematically in Fig.~\ref{fig:hidval}.
V-hadron production in Higgs boson decays was considered in
\cite{higgs}.  Here, we will consider a different scenario, in which
the v-hadrons are produced in LSsP decays.  In particular, as
illustrated schematically in Fig.~\ref{fig:valleySUSY}, production of
SM superpartners leads, through cascade decays, to the appearance in
the final state of two LSsP's.  If the LSvP is lighter than the LSsP,
then the LSsP will typically decay to an LSvP plus one or more
v-hadrons, some of which in turn decay visibly.  For simplicity we
assume in this paper both that R-parity is conserved 
and that the LSvP itself is stable; if either is violated,
the phenomenology may be richer still.


\begin{figure}[htbp]
  \begin{center}
    \leavevmode
     \epsfxsize=.45 \textwidth
     \hskip 0in \epsfbox{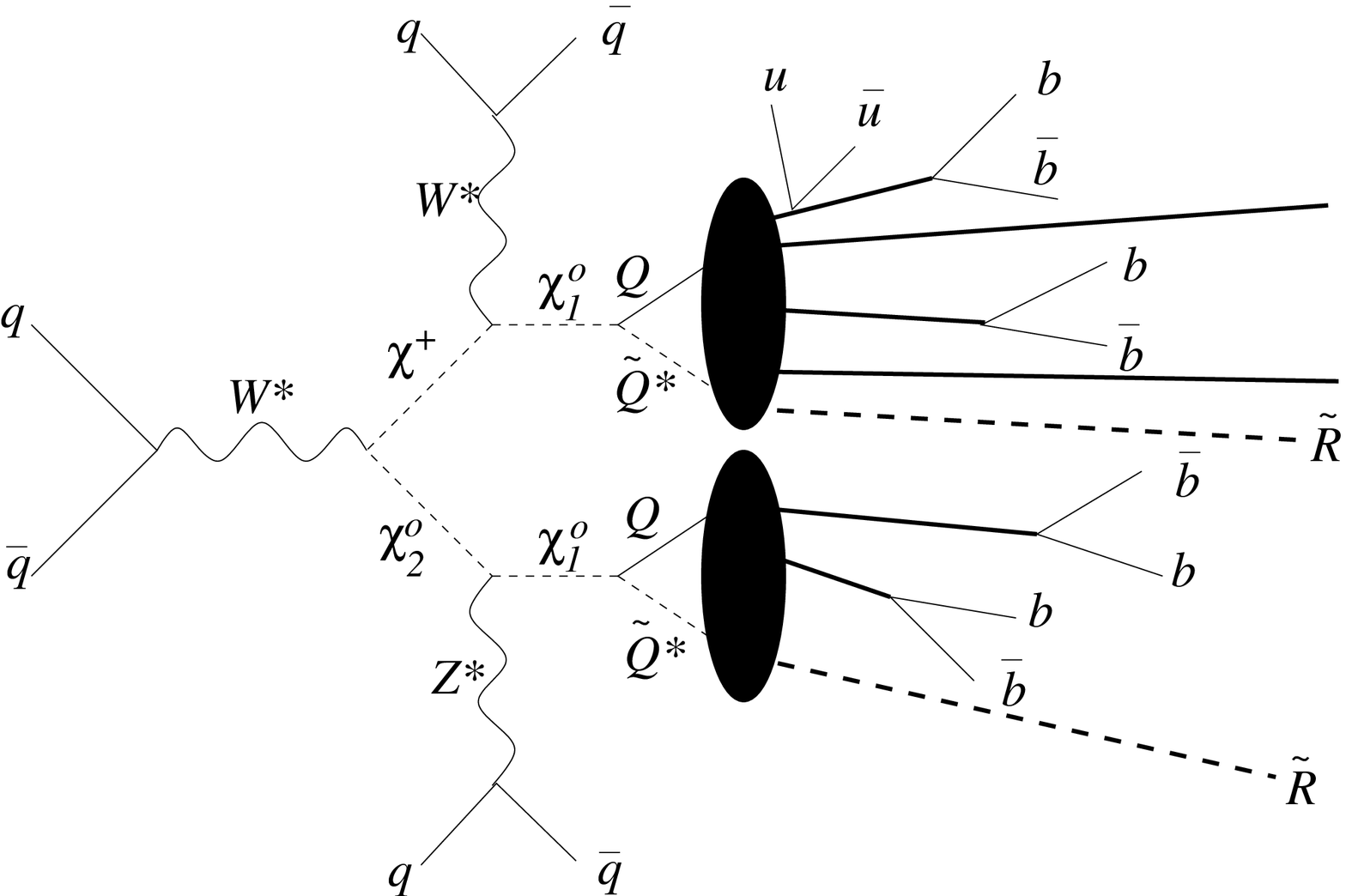}
  \end{center}
\caption{The production and subsequent decay of a chargino and
neutralino, showing the two LSsPs decaying to various v-hadrons, some
of which decay visibly.  Invisible R-parity-even (-odd) v-hadrons,
are shown as solid (dashed) lines; in particular, an LSvP,
labelled $\tilde R$, is produced in each of the LSsP decays.
}
\label{fig:susyevent}
\end{figure}

Let us now consider how phenomenology of LSsP decays in hidden-valley
models may differ in some ways from LSsP decays in other models.
First, since the LSvP is a v-hadron, its decay to the LSvP may be
accompanied by one or more long-lived R-parity-even v-hadrons,
possibly with a substantial multiplicity.  Some or all of these
v-hadrons may in turn decay to visible (but often rather soft)
particles.  This decay pattern may make the decay products of the LSsP
challenging to identify.  An example of how this could occur in SM
chargino-neutralino production is shown in Fig.~\ref{fig:susyevent}.
The two LSsP's ($\chi^0_1$) decay to a v-quark $Q$ and a v-squark
$\tilde Q^*$; after hadronization, a number of R-parity-even
v-hadrons and two R-parity-odd LSvP's ($\tilde R$) emerge.  Some of the
R-parity-even v-hadrons then decay to visible particles, leading to a
busy and complex event.  Second, many different v-hadronic final
states may appear in LSsP decays, just as a large number of QCD
hadronic states appear in $\tau$ and $B$ decays.  Acquisition
of a large sample of events may therefore require a combination of
search strategies.  Finally, since the LSsP and/or some of the v-hadrons it
produces may be long-lived and decay with highly displaced vertices,
discovery and study of these events may require specialized,
non-standard experimental techniques.

The reverse situation --- where the LSvP is
heavier than the LSsP --- is typically less dramatic,
but still worthy of note.  It leaves the bulk of SM SUSY signals
unchanged, but can in some cases produce spectacular and challenging
signals of its own.  It will be discussed briefly below.

Meanwhile, analogous statements apply, with only a few adjustments, in
other models with a conserved $\ZZ_2$ symmetry, such as
extra-dimensional models (or their deconstructed cousins) with a
spatial reflection symmetry in the internal dimensions (examples of
which \cite{universal} inspired the work of \cite{confused}.) The
process in Fig.~\ref{fig:susyevent} requires only minor modification;
for instance, the chargino and neutralinos might be replaced with
parity-odd Kaluza-Klein excitations of the $W$ and $Z$.  Little-Higgs
models with T-parity \cite{Tparity} are also potentially
affected. Indeed, the essential point generalizes further, to include
any conservation laws which are present and non-trivially
realized in both sectors.

\subsection{Hidden valley models}

What is a hidden valley and why is it so-named?  A hidden valley
sector (``v-sector'') has the following properties, illustrated in
Fig.~\ref{fig:hidval}.  First, like an ordinary hidden sector, it has
its own gauge symmetries and matter particles, with the property that
no particles (or at least no light particles) carry charges under both
standard model gauge groups and under the v-sector gauge groups.
Second, it has a mass gap, or at least a subsector with a mass gap, so
that not all v-particles can decay down to extremely light invisible
states which are stable or unobservably metastable.  Third, an
energetic barrier (a ``mountain'') limits cross-sector interactions
between the SM sector and the v-sector; this barrier prevented v-particle
production at LEP.  Fourth, low points through the mountain (the
elegant term ``portals'' was introduced recently in \cite{wilczek})
allow collisions of standard model particles at higher energy to
produce v-sector particles.  Fifth, massive long-lived v-sector
particles can decay back to light standard model particles by
tunnelling back through the mountains.  These decays have strongly
suppressed rates but are typically observable, sometimes with
displaced vertices.  And sixth --- a fact we will exploit here ---
cross-sector interactions may also allow for the Higgs boson itself
\cite{higgs}, and other neutral colorless SM particles such as the LSsP, to
decay into the v-sector (Fig.~\ref{fig:valleySUSY}.)

Since our main focus will be on models with an LSvP lighter than the LSsP,
it is important to emphasize that such models
are in no way unnatural. 
The mass spectrum of each sector depends both on
dynamics within that sector and on the mechanism by which it learns
about SUSY breaking.  There simply is
no reason for prejudice as to whether the two sectors have SUSY
breaking at the same scale, or whether one has larger breaking than
the other.  For instance, suppose SUSY breaking is communicated
strongly to one of the sectors but not the other.  Since the
interaction between the two sectors is suppressed, large SUSY breaking
in one sector directly induces only moderate or small SUSY breaking in
the other.  Even if both sectors learn about SUSY breaking with
comparable strength, it is a matter of detail whether the LSvP is
heavier or lighter than the LSsP.  Light v-gauginos are also easily
obtained in many models (indeed it can be a model-building challenge
to ensure they are not light) so this fact can easily lead to a light
v-gluino.

As in \cite{hidval}, we will consider hidden
valley models whose gauge groups are strongly-interacting and
confining, with the mass gap generated by strong interactions or by a
combination of strong interactions and the Higgs phenomenon.  The
strong interactions cause the v-sector particles to confine (at the
scale $\Lambda_v$) and form v-hadrons.  A number of long-lived
resonances will result.  The strong-interactions also cause v-parton
showering, following which, when the energy scale of a process is large
enough compared to $\Lambda_v$,
large numbers of v-hadrons may be
simultaneously produced.

A vast array of v-models are possible, with different details. For
definiteness, let us take the minimal supersymmetric standard model
(the MSSM) coupled via a $Z'$ (of coupling $\alpha'\equiv
g'^2/4\pi$ and
mass $M'=2 g' \hat v$) to a
v-sector of the form considered in \cite{hidval}, properly
supersymmetrized.  The v-sector will have v-quarks and v-gluons,
and their superpartners, transforming under an $SU(n_v)$ gauge group,
with coupling $\alpha_v\equiv g_v^2/4\pi$.
The simplest models require two Higgs-like chiral
multiplets $\phi_1,\phi_2$ in the v-sector, of opposite $U(1)$ charge
under the $Z'$, in order that anomalies cancel.  As usual,
$\hat v = \sqrt2\sqrt{\langle\phi_1\rangle^2+\langle\phi_2\rangle^2}$,
and $\tan\beta' =\langle\phi_2\rangle/\langle\phi_1\rangle$.
Higgs mixing, along
the lines of \cite{Wells,twin}, will lead to additional mixing
between the two sectors \cite{hidval, higgs}.  
Unfortunately, the scenario suffers from strong irreducible
model-dependence.  First, there are a minimum of six Higgs-like
electrically-neutral scalars (the usual three in the SM sector and
three more in the v-sector, of which one in each sector is naturally
CP-odd) and seven neutralinos (the usual four plus $\tilde Z',
\tilde\phi_1, \tilde \phi_2$ in the v-sector.)  Mixings among these
particles can take many forms.  Second, as in GMSB and
RPV models, an unstable LSsP 
need not be neutral or colorless; thus there are many possibilities
to consider.
Finally, there are many choices for the properties of the $Z'$.  A full
study of parameter space is not feasible, and would not be particularly
useful, given that this is just one of many models.
Instead, an attempt will be made here to point out contexts
where the phenomena are distinct from those in
GMSB, RPV and other models with an unstable LSsP, and give 
signals that are interesting and often difficult (but probably
not impossible) for the Tevatron and/or LHC experiments to detct.

\subsection{Ordinary v-hadrons and their decays}

\begin{figure}[htbp]
  \begin{center}
    \leavevmode
     \epsfxsize=.45 \textwidth
     \hskip 0in \epsfbox{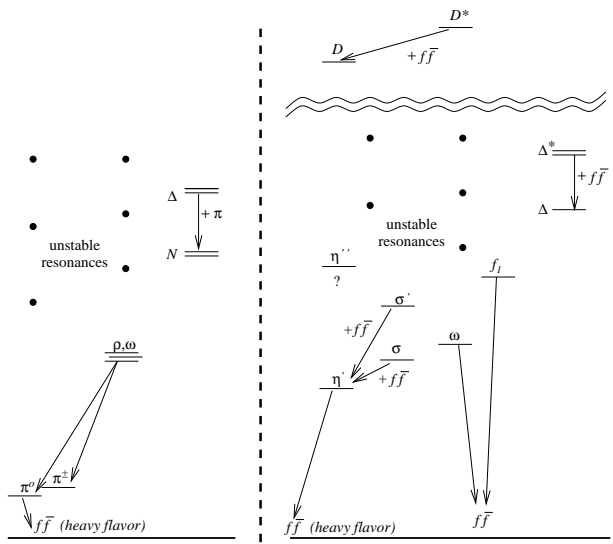}
  \end{center}
\caption{Partial
spectrum and decay modes in the two-light-flavor regime (left)
and one-light-flavor regime (right); the latter is partly guesswork.}
  \label{fig:12LFspect}
\end{figure}

Since every supersymmetric production event leads to
production of v-hadrons, we start with a discussion
of their spectrum, lifetimes and decay products.  The spectrum of a
v-sector consisting of a QCD-like theory with two v-quarks, one or
both of which are light, was discussed in \cite{hidval}, and
is reproduced in Fig.~\ref{fig:12LFspect}.  If both v-quarks are light ---
the two-light-flavor (2LF) regime --- compared to the confinement
scale $\Lambda_v$, the
spectrum is familiar from the real world, with all v-hadrons decaying
quickly to a v-isospin triplet of light v-pions or to v-baryons.  If
one v-quark is light (the 1LF regime), the spectrum is not fully known;
several v-hadrons with comparable masses but different
spins and CP-quantum numbers are expected to be stable against decay
to other v-hadrons.  In both cases, the decay lifetimes \cite{hidval}
are such that displaced vertices are possible if the v-hadrons
are light, and/or the $Z'$ that couples the two sectors is heavy, 
and/or a flavor-changing neutral current is roughly
of the right size.  

Some reanalysis of the results of \cite{hidval} is necessary for the
present context.  In a model with sufficiently large Higgs mixing and,
as is the case here, both CP-even and CP-odd Higgs scalars in both
sectors, Higgs-mediated decays can compete with or exceed those
mediated by the $Z'$ discussed in \cite{hidval}.  Note also that it
was assumed in \cite{hidval} that the $Z'$ is light enough to be
produced at the LHC; here, v-hadron production proceeds via a
different mechanism, so there is no such constraint.  This widens the
range of possible v-hadron lifetimes and increases the likelihood that
the Higgs sector will dominate v-hadron decays.  The discussion will
be brief, since so many other models can be written in which similar
or novel mechanisms for long-lived particles can be obtained.  The
important points to emphasize will be these: (1) the decays need not
be unobservably slow; (2) the decays need not be prompt; (3) heavy
flavor is likely to be common in the final state; (4) decays to
$\tau^+\tau^-$ are expected, and branching fractions to $\mu^+\mu^-$
pairs may be measurable.  To keep this already broad discussion under
control, we will assume a relatively light mass spectrum, for which
decays to $t\bar t$, $WW$ and $ZZ$ are unimportant or simply absent.
If they are present, some additional experimental issues arise.

If we assume that mass splittings from SUSY breaking in the v-sector
are large compared to $\Lambda_v$, then the lightest v-hadrons are the
R-parity-even states shown in Fig.~1 of \cite{hidval}.  The decays of
these v-hadrons to the SM-sector can occur through the v-Higgs sector
(whose states we will simply refer to as $\phi$) or via the $Z'$.
Although the Higgs sector may have a light $\phi$, the $Z'$-mediated
decay can be stronger than the $\phi$-mediated decay if $\hat v$ is
not too large, and/or Higgs mixing is not too large, and/or the Higgs
Yukawa coupling to the v-quark and v-squark is not large.  For
definiteness, let us consider a simple case.  To see whether the $Z'$-
or $\phi$-mediated decay is most important, we estimate both processes
neglecting the generally-subleading effect of interference.  In the
2LF regime, the most interesting particle is the $\pi_v^0$.  Its decay
via a $Z'$ to $b\bar b$ was computed in \cite{hidval}, where
a substantial suppression for small $\pi_v^0$ masses, and a
resonant enhancement for $m_{\pi_v^0}\sim m_Z$, was noted:
\begin{equation}\label{Gpi2ffZ}
\Gamma_{\checker\pi^0}
={3\over 32\pi}{1\over \hat v^4}{
Q_{H}^2 f_{\checker\pi}^2 m_{\checker\pi}^5
\over (m_{\checker\pi}^2-m_Z^2)^2}
m_b^2  \ .
\end{equation}
For $4 m_b^2 \ll m^2_{\pi_v}\ll m_Z^2$,
\begin{equation} \\
\Gamma_{\checker\pi^0} \sim
6\times 10^{9} {\rm \ sec}^{-1}{f_{\checker\pi}^2 m_{\checker\pi}^5\over
(20 \GeV)^7}
\left({5 \TeV\over  \hat v}\right)^4  
\ .
\end{equation}
Here $Q_H$ is the charge of 
$H$ under the $Z'$, taken to be $2/5$ for
the numerical estimate.
In comparing to \cite{hidval} we have used $m_Z' = 2g'\hat v$ in
this model.  
For $m_{\checker\pi}\ll m_Z$ and $\hat v\agt 3$ TeV, this is
macroscopic.
Note $\hat v$ in \cite{hidval} was taken
of order 5--10 TeV to allow reasonable production rates of v-quarks
at the LHC, but here $\hat v$ could be much larger, as large
as 100 TeV or even more, without losing the v-hadron phenomenology.

An estimate
of the $\pi_v^0$ decay via a CP-odd scalar of mass $m_A$
can be obtained as follows. (A full
calculation is straightforward but beyond our present needs.)
If the lightest CP-odd scalar is much heavier than the $\pi_v^0$ and
if its component in the SM-sector (v-sector) is $\cos\theta_-$ 
($\sin\theta_-$), 
then, defining
$\tan \beta\equiv {v_u/ v_d}\equiv {\vev{H_u}/ \vev{H_d}}$,
we have
\begin{equation}\label{Gpi2ffA}
\Gamma_{\checker\pi^0}
= {3\over 8\pi}{\theta_-^2\over \hat v^2 v^2}
{ f_{\checker\pi}^2 m_{\checker\pi}^5 \over (m_{\checker\pi}^2-m_A^2)^2 }
m_b^2 \ f(\beta') \ \tan^2\beta
\end{equation}
The function $f(\beta')$
could most naturally be either $\tan^2\beta'$ or $\cot^2\beta'$, depending
on the coupling of $A$ to the light v-quarks.
In this expression $\theta_-$, the mixing parameter between the CP-odd
scalars in the two sectors, is assumed small.  Its 
natural size (if it is induced by D-terms  
as in \cite{Wells} ---
though this can be evaded in nonminimal models with F-terms, as was used
in  \cite{higgs})
is of order $2 Q_H v/\hat v$.  Consequently
both (\ref{Gpi2ffZ}) and (\ref{Gpi2ffA}) 
naturally scale as  $\hat v^4$.  
The two processes are parametrically of the same order;
details of the spectrum and mixing angles determine
which is larger. For
any $m_{\pi_v}$ (except very near the $Z$ or $A$ masses)
there exist ranges of 
reasonable 
$\hat v$ for which the $\pi_v^0$ decay is respectively
prompt, displaced, or external to the
detector.  
The $\pi_v^0$ branching
fractions are similar to those of a CP-odd Higgs boson of the same
mass; in the decay via a $Z'$, this is due to helicity suppression
\cite{hidval}.  The $\pi_v^\pm$ may be stable, in which case it
contributes only MET, but in some models the $\pi_v^\pm$
will decay through a flavor-changing coupling.  In this case its
lifetime is longer than that of the $\pi_v^0$, by an amount determined
by the strength of this coupling.

In the 1LF regime, the $\eta'_v$ decay is similar to
that just described for the $\pi_v^0$.  The $\omega_v$ decays
predominantly via an off-shell $Z'$ to $f\bar f$ ($f$ any SM fermion) 
with branching fractions determined by the $Z$ and $Z'$ \cite{hidval}.
Its decay is much faster
than that of the $\eta'_v$; roughly,
\begin{equation}
\Gamma_{\checker\omega}
\sim {10^{18} \ {\rm sec}^{-1}}\left({m_{\checker\omega}^5\over 200\ {\rm GeV}}\right)^5
\left({5\ {\rm TeV}\over \hat v}\right)^4 \ .
\end{equation}
%
The decay $\omega_v\to\eta'_vf\bar f$ is
generally suppressed unless mass splittings
permit an on-shell decay to $\eta'Z$ or $\eta'h$.

As for the lightest R-parity-odd v-hadron, it generally consists of a
(typically heavy) v-squark or v-gluino bound to one or two v-quarks or
v-gluons.  Such a heavy-light system is well studied in the context of
SM heavy-quark mesons.  
elsewhere.)  The lightest R-parity-odd v-hadron ``$\tilde R$'', the
LSvP, is stable against v-hadron decays, and may be truly stable if it
is the LSP.  Most other R-parity-odd v-hadrons have v-strong decays
$\tilde R$ plus other v-hadrons.  However, the spin splitting between
the LSvP $\tilde R$ and its first excitation $\tilde R^*$, which decreases as
$m_{\tilde R}/\Lambda_v$ increases, may be small enough to stabilize
the latter against v-hadronic decays.  The decay $\tilde R^*\to f\bar
f \tilde R$, which may occur via an off-shell Higgs or $Z'$ boson, may
be experimentally interesting.


\begin{figure}[htbp]
  \begin{center}
    \leavevmode
     \epsfxsize=.45 \textwidth
     \hskip 0in \epsfbox{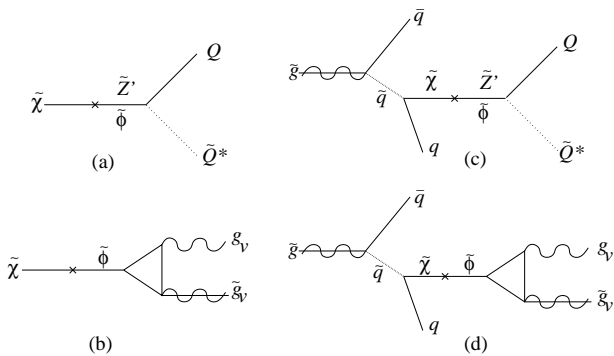}
  \end{center}
\caption{Various possible decays of an LSsP to an LSvP.
(a) At one extreme, prompt neutralino decays to a v-quark and v-squark
via mixing of the neutralino with the $\tilde Z'$ or $\tilde \phi$.
(b) Slow decay of a neutralino to a v-gluon and v-gluino via a 
v-quark/v-squark loop.  (c) Slow decay of a gluino to $q\bar q$ plus
a v-quark and v-squark.  (d) Very slow decay of
a gluino to a v-gluino.}
  \label{fig:LSsPdecays}
\end{figure}

\subsection{LSsP decay rates}

We now turn to the decay of the LSsP to the v-sector.  A complete
analysis of the possibilities is beyond the scope of this article.
Here the aim is to consider generally the signals that might arise, in
order to initiate discussions of the experimental
implications and to
motivate future studies.

Let us first see that the LSsP's decay may be prompt, unobservably
long, or displaced.  Initially  we will focus on the scenario
where $M$ and $m$, the masses of the LSsP and LSvP, satisfy
$M-m\gg \Lambda_v$; we will deal briefly with
other cases later.  The LSsP decay rate depends of course on the precise
v-spectrum and the composition of the LSsP and LSvP, and relevant
mixing angles can vary widely, so we content ourselves with rough
estimates here.  For obtaining figures of merit, we will take typical
mixing angles between the SM-sector neutralinos and the v-sector
neutralinos to be of order $\theta_v\sim v/\hat v$, which
is typically $\sim 0.05$; this
assumption will not be crucial to the conclusions.  

If the LSsP is a neutralino, then as shown in
Figs.~\ref{fig:LSsPdecays}a and \ref{fig:LSsPdecays}b, its decay can
occur either through a $\tilde Z'$ or $\tilde \phi$ coupling to
v-quark and v-squark $Q\tilde Q^*$, or through a $\tilde \phi$
coupling to a v-gluon and v-gluino $g_v\tilde g_v$.  If the $\tilde
Z'$ dominates, then the decay rate is of order
\begin{eqnarray}\label{chidecay}
\Gamma_{\tilde \chi}&\alt& {\alpha' m_{\tilde\chi}n_v} \theta_v^2 \nonumber
\\
&\sim& 2\times 10^{22} \ {\rm sec}^{-1} \
\left({\alpha'\over 0.01}\right)\left({m_{\tilde\chi}\over 200 
\GeV}\right)
\nonumber \\
& & \times
\left({\theta_v\over v/\hat v}\right)^2 
\left({5 \TeV \over \hat v}\right)^2  \ . 
\end{eqnarray}
(In the last step we took $n_v$, the number of v-colors, to be 3.)
This decay is prompt, even accounting for possible additional
phase-space suppression.  If instead an intermediate $\tilde \phi$
dominates through couplings to $Q\tilde Q^*$, the coupling $\alpha'$
in (\ref{chidecay}) should be replaced by a coupling $y_Q^2/4\pi$,
where $y_Q$ is the coupling of the relevant v-quark to the relevant
v-Higgs, but the decay remains prompt.  If there are no light
v-squarks, a neutralino decay to a v-gluino can occur through a loop,
as in Fig.~\ref{fig:LSsPdecays}b, and is of order
\begin{eqnarray}\label{chidecay2}
\Gamma_{\tilde \chi}&\alt& {n_v^2-1\over 4\pi}{m_{\tilde\chi}} \theta_v^2 
\left({\alpha_v n_v n_H m_{\tilde\chi} \over 48\pi\hat v}\right)^2 \ 
\nonumber \\
&\sim &
2\times 10^{16} \ {\rm sec}^{-1}
\left({m_{\tilde\chi}\over 200 
\GeV}\right)^3
\nonumber \\
& & \times
\left({\theta_v\over v/\hat v}\right)^2 
\left({5 \TeV \over \hat v}\right)^4  \ .
\end{eqnarray}
Here $\alpha_v$ and $n_v$ are the coupling and number of colors of the
v-confining gauge group, and $n_H$ is the number of heavy v-quarks
obtaining masses through $\phi$; in the above estimate we took $n_v=3$
and 
$\alpha_vn_vn_H\sim 1$.  We have assumed v-squark masses are
parametrically of order $\hat v$; if they are much larger they will
further suppress the amplitude.  In some regions of parameter
space, this decay can occur with a displaced vertex, though it is
often prompt.

Suppose instead the LSsP is a gluino. Then it may decay to a v-quark
and a v-squark plus either a quark-antiquark pair (as shown in
Figs.~\ref{fig:LSsPdecays}c) or a gluon (not shown, and typically with
a smaller branching fraction).  The rate for the former, if dominated
by gaugino exchange, is of order or smaller than
\begin{eqnarray}\label{chidecay3}
\Gamma_{\tilde g}&\alt& {\alpha_s\alpha_2\alpha' N_f n_v m^7_{\tilde g}\over
m_{\tilde q}^4m_{\tilde\chi}^2} \theta_v^2 \nonumber \\ 
&\sim& 
6\times 10^{15} {\rm sec}^{-1}\ 
\left({\alpha'\over 0.01}\right)
\left({m_{\tilde g}\over 500 
\GeV}\right)^7 \left({\theta_v\over v/\hat v}\right)^2
\nonumber \\
& & \times
\left({1 \TeV \over {m_{\tilde q}}}\right)^4
\left({1 \TeV \over {m_{\tilde \chi}}}\right)^2
\left({5 \TeV \over \hat v}\right)^2  \ .  
\end{eqnarray}
Here $\alpha_2= g_2^2/4\pi$ is the coupling of the SM $SU(2)$ group
factor, and $N_f$ is the number of squarks of mass $m_{\tilde q}$; we
took $N_f=1$ and $n_v=3$ for this estimate.  This estimate is very
rough and subject to large parameter dependence and phase space
suppression.  The direct coupling of the $\tilde Z'$ to the
intermediate-state SM squark $\tilde q$ may well be more important; the
rate is as above with the replacement
$\alpha_2\theta_v^2/m_\chi^2$ with $Q_q^2/\hat v^2$, where $Q_q$ is
the charge of $\tilde q$ under the $Z'$ \cite{hidval}.  A similar
formula governs if higgsino-exchange is dominant, with a possible
Yukawa-coupling (or v-coupling) suppression.  This decay mode may be
prompt or may produce a displaced vertex within the detector.  Since
the gluino is long-lived compared to the QCD scale, it will undergo
hadronization, and its phenomenology is similar to that discussed in
\cite{split}.  If no v-squarks are kinematically available, the gluino
will decay to a v-gluino and a v-gluon, as shown in
Fig.~\ref{fig:LSsPdecays}d.  This is a very slow decay, and will
generally occur outside the detector --- or, if the hadron containing
the gluino stops within the detector \cite{split}, it will have an
out-of-time decay, such as searched for in \cite{D0split}.  The final
state, containing v-hadrons as well as a quark and antiquark,
differs somewhat from that considered in \cite{split, D0split}, but is
presumably still constrained.

Finally, decays of LSsP sleptons or squarks are intermediate between
these possibilities.  The decays are typically (but not always) prompt.
A lepton or quark accompanies the v-hadrons in each LSsP decay.  A variety
of phenomena are possible; we will not give an exhaustive list here.

Let us summarize the phenomenological possibilities and reemphasize
the main points.  Neutralinos can directly link the two sectors and
decay promptly.  Charged matter (LSsP squarks, sleptons or charginos)
must first emit neutralinos to cross the barrier between the sectors.
Gluinos and v-gluinos must first couple to charged matter, which in
turn couples to neutralinos, in order to communicate between the two
sectors.  The more steps required to cross from one sector to the
other, the slower the LSsP decay rate.  The multiple possible choices
for LSsP and LSvP allow great variability in the LSsP lifetime.  Also,
as in GMSB models, a charged or colored LSsP must decay to a standard
model particle in addition to v-hadrons, enriching the
signals.
 
It is important to emphasize that in contrast to \cite{hidval, higgs},
where it was vital that the scale $\hat v$ of the ``portal'' between
the two sectors not be much above 10 TeV, in order that v-hadrons be
produced with measurable rates, the present scenario has a much weaker
requirement, especially in the 1LF regime.  The above formulas for
LSsP and v-hadron decay rates show that values of $\hat v$ in the 100
TeV range, or even higher, may still give detectable decays.  If this
turns out to be the case, LSsP decays will be the only source of
v-hadron production, and the only clue to physics both in the v-sector
and at the 100 TeV scale.

\subsection{Final states of LSsP decays}

The final state of the LSsP decay is affected by the v-strong
interactions; it contains one $\tilde R$, plus other v-hadrons
(including possibly an even number of additional R-parity-odd states.)
In the 2LF regime of the theory studied in \cite{hidval}, the LSsP
typically decays to one $\tilde R$ and $n$ $\tilde \pi_v$, where $n$
will fluctuate from event to event.  (For instance, if $m_{\tilde\chi}
\sim 200$
GeV, $m_{\tilde R}\sim 60$ GeV, $m_{\pi_v}\sim 20$ GeV, experience
from SM $\tau$ decays would suggest $n\sim$ 0--4.)  
The $\pi_v^0$ will decay visibly; whether the $\pi_v^\pm$
decays visibly is model-dependent.
The 1LF regime of the same theory probably will have smaller $n$ for
the same $\Lambda_v$, but has fewer invisible v-hadrons, and possibly
a larger average number of visible jets and leptons per v-hadron
\cite{hidval}.  The final state of each LSsP will consist of MET and
up to $\sim 2n$ moderately soft and poorly isolated objects, with a
preponderance of heavy-flavor quarks and $\tau$s.  If all v-hadron
decays are prompt, displaced vertices from the many $b$ mesons may be
notable, but the low momentum of the $b$ quarks and their overlapping
jets may make $b$-tagging less than maximally efficient.  If some
v-hadrons decay late, there may be as many as $n$ highly displaced
vertices for each LSsP.

In some kinematic regimes, the signals may be similar to those
familiar from other models, such as those with GMSB.  For instance, if
$M-m\sim \Lambda_v$ (where $M$ and $m$ are the masses of
the LSsP and LSvP,) the predominant decay of the neutralino LSsP
may be $\tilde\chi\to \tilde R\pi_v^0$, that is, $n=1$.
If $m_{\pi_v}$ or $f_{\pi_v}$ are too large,
the decays $\tilde \chi\to Z\tilde R$ 
or $\tilde \chi\to h \tilde R$
may be important, or,  
if even this is
kinematically forbidden, then three-body decays such as $\tilde\chi\to f\bar
f \tilde R$ may result.  In such contexts the final state of each LSsP
still has ample MET, along with a SM fermion pair that accompanies
every (or almost every) LSsP decay.  With luck, this may be a
relatively hard $b\bar b$ or lepton pair.  Then
every supersymmetric event has four moderately hard fermions, and the
MET signal is reduced by a relatively small factor, so standard SUSY
searches with a demand for leptons or extra $b$-tags may be able to
isolate a signal.  But the larger is $M-m$ relative to $\Lambda_v$,
the larger is $n$, giving an unfamiliar final state with reduced MET,
many soft jets and/or leptons, and possibly many displaced vertices.

The degree to which the MET signal is reduced is model-dependent.  
A very crude estimate is simply as follows.  If $m,\Lambda_v\ll M$,
and $k$ of the $n$ v-hadrons are invisible, then on average the MET
signal is reduced, relative to the same event with the LSsP escaping
the detector, by a factor of order $(k+1)/(n+1)$.  For $\Lambda_v\ll
m\sim M$, the reduction factor approaches $m/M + [k/n](1-m/M)$.  The 
largest reduction in MET thus occurs for $m\ll M$,
$n$ large (requiring $M-m\gg \Lambda_v$), 
and $k\ll n$ (requiring most v-hadrons decay visibly).  
For instance, the 1LF regime with $M\gg m\gg
\Lambda$ would exhibit a very strong reduction of the MET signal, as
would the 2LF regime if the $\pi^\pm$ decays visibly. Conversely,
a substantial
reduction can only be avoided if most v-hadrons decay invisibly or if
$m\sim M$.
Since SM backgrounds fall steeply with
MET, any appreciable reduction in the MET signal greatly
decreases the signal-to-background ratio throughout the MET
distribution.
Without the strong MET signal, and in the absence of easily
identifiable ``tagging'' objects such as photons, novel search
strategies may be needed to discover supersymmetric particle
production.

\subsection{Production and detection}

At the Tevatron, the number of supersymmetric events in models not
excluded by LEP could be as large as $100$ to $1000$ in current data.
In the hidden-valley scenario, the detection of these events could be
made much more subtle --- though possibly easier, in the end --- than
expected.  Consider again Fig.~\ref{fig:susyevent}; would any analysis
so far have detected such phenomena?  On the one hand, it is not
obvious that a hundred events with 4 to 8 soft $b$ jets and relatively
low MET, possibly with additional hard jets, possibly without, would
yet have been noticed.  The trigger efficiency on such events,
assuming a trigger on muons, jets and/or MET, might well be of order
20-40 percent, so a substantial sample would have been recorded, but
no typical search strategy might have isolated them as yet.  On the
other hand, since these events may have multiple highly displaced
vertices, more searches for such vertices might expose these events
directly.

At the LHC, the number of supersymmetric events produced may be
enormous, perhaps $10^3$--$10^5$ each year.  The number of v-hadrons
produced in each event may be of order $2$--$10$.  With $10^3$--$10^6$
v-hadrons, branching fractions of a percent or less become
experimentally interesting.  Both triggering and event-selection may
benefit from the many $\tau$ pairs and $\mu$ pairs (either produced
directly or in $b$ and $\tau$ decays) that will be present in these
events.  If a clean and sufficiently large sample of such events is
obtained, perhaps through highly-displaced vertices from LSsP or
v-hadron decays, it may even be possible to do some precision
spectroscopy using lepton-pair invariant mass distributions, looking
for resonances or kinematic endpoints that could arise from
$\omega_v\to\mu^+\mu^-$, $\tilde R^*\to \mu^+\mu^- \tilde R$, etc.
(Note there will be an irreducible combinatoric background from
leptons arising in v-hadron decays to $b$'s and $\tau$, so high
statistics and wise selection criteria will be important to find such
kinematic features.) Conversely, in scenarios where the number of
supersymmetric events is small, the highly-displaced vertices may be
the key experimental signal by which supersymmetric particles (as well
as, simultaneously, the v-sector) are first discovered, and clean
samples are obtained.

Because both the LSsP and the v-hadrons in its decay can be
long-lived, with various possible lifetimes, a wide variety
of track patterns must be searched for.  The LSsP decay may produce, for
example, (i) a single vertex with a very large number of tracks, or
(ii) a set of displaced vertices with multiple tracks that themselves
all point back to a single displaced point with very few or no tracks,
or (iii) a displaced vertex with many tracks, followed by other
vertices with fewer tracks.  Other possibilties abound.  Many of these
tracks or clusters of tracks may not point back, even approximately, 
toward the beampipe,
making them hard to identify.  It may be worth considering how best to
design tracking algorithms to recognize candidates for such decays, so
as to allow events to be flagged for further off-line analysis.  The
challenge may be significant, since many of these events may be very
busy, with the various displaced vertices occuring within clusters of
tracks emerging from overlapping soft jets.


Looking for highly displaced vertices (and/or for many
slightly-displaced vertices) will be difficult and may be
computationally expensive.  It is therefore important to devise
effective strategies for selecting events that are especially likely
to contain them.  It has been suggested that high-precision timing
information may serve to identify events that contain
highly-displaced vertices \cite{timing}; while this study focuses on
late-arriving photons, it presumably also applies to hard electrons and
to $pi_0$ decays to photons.  Other approaches might include the
following:
\begin{itemize}
\item
A sample of events with moderate
MET (with or without high-$p_T$ jets) could be searched
both for an unusual number of displaced vertices inside the beampipe,
and for vertices outside the beampipe.  
\item
Events that already have three
or more displaced vertices detected within the beampipe would be good
candidates for additional vertices outside the pipe.  
\item
Events with multiple {\it and not necessarily isolated} muons would
form a good sample in which to check for a $\mu$ pair emanating from a
out-of-pipe vertex.  The vertex might have several other tracks
(expected in $b\bar b$) or be non-pointing (expected in
$\tau^+\tau^-$) or, if there are no other tracks, have an invariant
mass far from any known resonance.  The scenarios covered would extend
somewhat beyond those affected by the search carried out in
\cite{mumu}, which focuses on cleaner signals with no background, and
has lower than maximal acceptance.
\item
Another interesting sample would include events with two or more
``trackless'' jets --- jets that lack stiff tracks pointing near the jet
axis.  Within such jets, one could look for muon candidates
registering in the muon chamber but which lack a corresponding track
in the innermost detector elements.  Such muon-candidates within
trackless jets would suggest the presence of a displaced vertex outside
the innermost tracker.
\item
Additional searches for displaced $Z$ or $h$ candidates, in channels
other than $Z\to e^+e^-$ searched for in \cite{cdfZsearch}, would be
welcome, perhaps using the negative-impact-parameter tracks method
discussed in \cite{LSsPtoZh,stopLSsP}.
\end{itemize}

If no readily-produced new particles decay with highly displaced
vertices, additional techniques may need to be considered.  Although
many supersymmetric events with a decaying LSsP may have 4 or more $b$
quarks and/or $\tau$s, the efficiency for tagging of $b$ quark jets
and $\tau$ identification in an environment such as that shown in
Fig.~\ref{fig:susyevent} is surely unknown, and probably low.  Also,
jet reconstruction in such a busy event will be problematic.  A
program of studies, along the lines of \cite{4tau}, but more extensive
and involving new Monte Carlo tools and full detector simulations,
will be needed to clarify these issues and allow for effective event
selection.

As an aside, let us note that, once a working strategy for event
selection has been established, it may be possible to test an
important prediction of this scenario, one which is also true of other
scenarios with LSsP decays.  Some fraction of events may produce two
highly boosted LSsP's, which may in turn produce collimated and
readily identifiable decay products.  These products roughly indicate
the momentum directions of the LSsP's.  Since the missing transverse
momentum in such events should come mainly from the LSvP's and other
invisible v-hadrons from the LSsP decay, the direction of the missing
momentum should lie between the two putative LSsP momentum directions.
This prediction is a test of the two-stage process that forms the
events: (1) cascade decays of a pair of SM supersymmetric particles
lead to two boosted LSsP's, and then (2) the decay of the two LSsP's
produces two collimated clusters of both visible and invisible
particles.

\subsection{A heavier LSvP}

Let us briefly consider the situation when the LSvP is heavier than
the LSsP.  In this case, the vast majority of events in which SM
superpartners are produced will not be affected by the v-sector.  Only
v-sector production, assuming it occurs at all, may be affected in
interesting ways.  In supersymmetric v-production, each
event will have two LSvP's, each decaying to an LSsP, but
the decay of the LSvP itself may not be striking.
For instance, a v-squark LSvP may simply decay (promptly) as $\tilde Q
\to Q \tilde\chi$, where $\tilde\chi$ is a neutralino LSsP.  This
decay occurs within a v-hadron $\tilde R$, so the final state
will often contain one or more v-hadrons.
But in this context, v-quark pair production and v-squark
pair production may not differ greatly.  In many scenarios, both 
classes of events will
produce multiple visible and invisible v-hadrons; the v-squark
events will have larger MET on average, but on an event-by-event basis
they may look very similar.

For $M/\Lambda_v$ not large, v-gluino pairs will be produced in parton
showers, just as light quarks and even charm quarks are pair-produced
in QCD.  Consequently the number of $\tilde R$ hadrons produced in an
event may well be larger than two.  In some cases (see below) this
could lead to striking signatures.  Note also that v-production
of R-parity-even states, such as v-quarks, 
may generate $\tilde R$ hadrons in the final state through the
parton showering.



Suppose now that $\tilde R$ and $\tilde\chi$, with mass $M$ and $m$,
lie close in mass, in particular with $M-m\alt\Lambda_v$.  This may
naturally occur if the bare gluino mass is small compared to
$\Lambda_v$.  Then the decay $\tilde R\to \tilde\chi$ plus v-hadrons
may be forbidden.  In this case decays such as $\tilde R \to Z\tilde
\chi$ and $\tilde R\to h \tilde \chi$ may occur; though mediated by a
loop-induced $g_v\tilde g_v \tilde \phi$ interaction, these will still
typically be prompt, as the crude estimate
\begin{eqnarray}
\Gamma_{\tilde R\to \tilde\chi Z}&\sim& 
\alpha_2\theta_v^2\left({\alpha_v n_v n_H\over 48\pi \hat v}\right)^2 
{\Lambda_v^3}\nonumber \\
&\agt& 2\times 10^{14} \ {\rm sec}^{-1} 
\left({\Lambda_v \over 300  \GeV}\right)^2
\nonumber \\  & & \times 
\left({5 \TeV \over \hat v}\right)^4
\left({\theta_v\over v/\hat v}\right)^2 
\end{eqnarray}
demonstrates.  
A decay with a displaced vertex becomes more
likely if $M-m\ll m_Z$, however, since in this case 
the rate receives an additional
suppression by $\sim \alpha_2(M-m)^4/m_Z^4$.
This phenomenon is not so different
from neutralino decays in some GMSB models, which may produce a
gravitino plus a $Z$ and $h$, possibly at displaced vertices
\cite{LSsPtoZh}; but photons are also generally produced as well
in such models, while that is unlikely here.

\begin{figure}[htbp]
  \begin{center}
    \leavevmode
     \epsfxsize=.45 \textwidth
     \hskip 0in \epsfbox{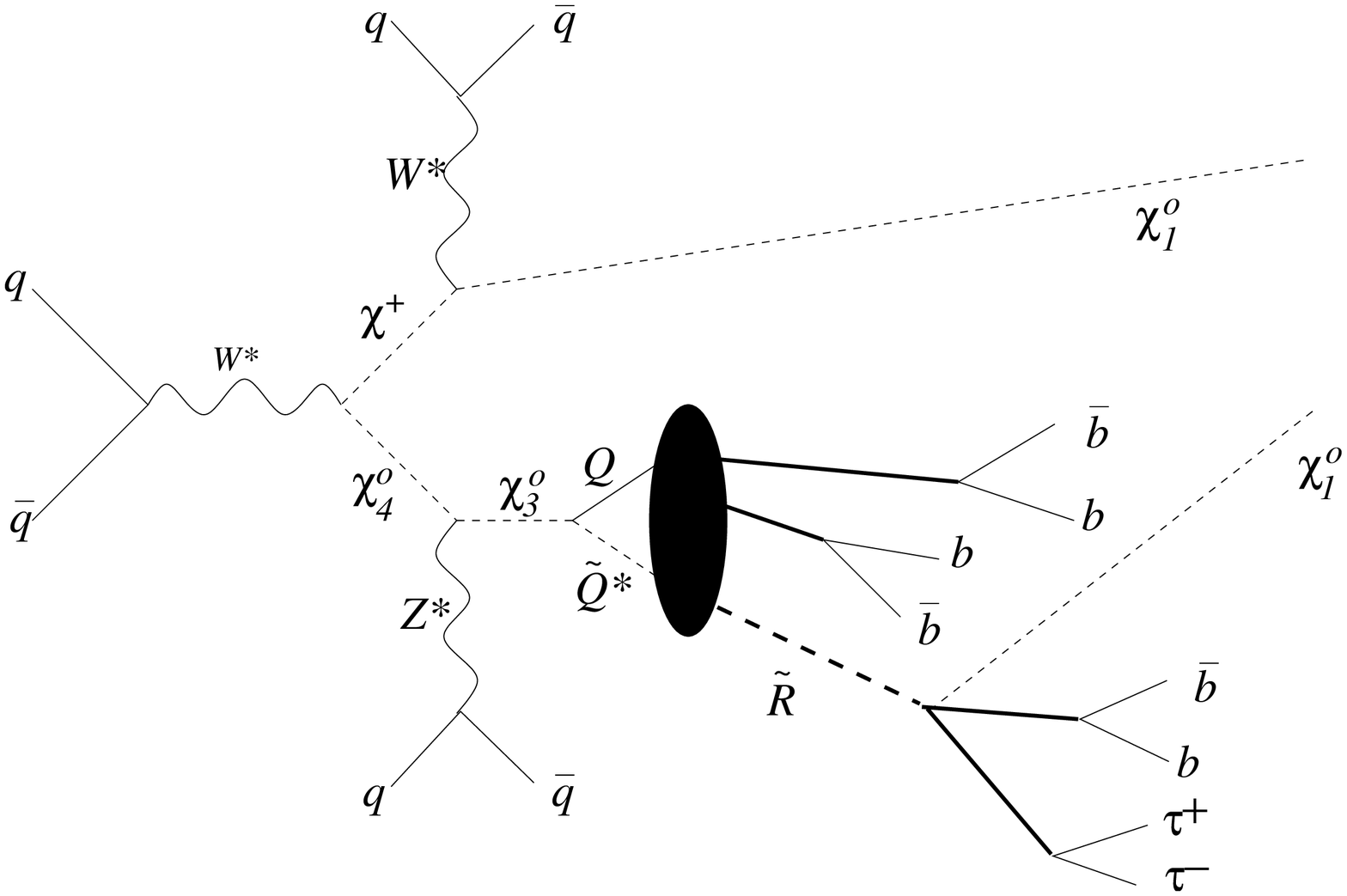}
  \end{center}
\caption{Cascade decays following
production of SM superpartners
may occasionally lead, through neutralino mixing, 
to v-hadron production.
}
\label{fig:crossportal}
\end{figure}

If $\hat v$ is very large, direct v-quark or v-squark production rates
via a $Z'$ may be too small to observe.  Even in this case, however,
SM superpartner production offers another opportunity to discover the
v-sector.  During the cascade decays following sparticle production,
mixing among neutralinos and/or Higgs bosons from the two sectors can
allow v-sector particles to be produced, as shown in
Fig.~\ref{fig:crossportal}.  While this will be a rare effect, the
large cross-section for SM sparticle production means that this
process can still be an important, or even the dominant, mechanism for
v-hadron production.

\subsection{Analogues in extra dimensional models, etc.}

Essentially all of the remarks above have analogues in, for example,
models with universal extra dimensions \cite{universal}, or indeed
dimensions of any sort, in which a $\ZZ_2$ symmetry (KK-parity) plays
a role similar to R-parity in both the SM and the v-sector. The
processes shown in Figs.~\ref{fig:susyevent}, \ref{fig:LSsPdecays} and
\ref{fig:crossportal} can occur in these models as well, with chargino
and neutralinos replaced with KK-excitations of charged and neutral
vector bosons, the $\tilde R$ with a v-hadron containing the lightest
$\ZZ_2$-odd KK-excitation of the v-gluon or a v-squark, {\it etc.}
The phenomenology will be different in detail, but will at first
experimental glance have similar features, in that the expected
missing energy signals are diluted, soft jets are produced, {\it
etc.}.  This is in line with the remarks of \cite{confused}.  Similar
issues would apply to little Higgs models with a T-parity
\cite{Tparity}, assuming T-parity also acts on the v-sector.

A difference between these theories and supersymmetry is that the
operators connecting the two sectors may be of higher dimension in the
present case.  For instance, squark decays to v-squarks may proceed
via dimension five operators, while KK-quarks decaying to KK-v-quarks
always do so through operators of dimension six.  Thus the lifetime of
the lightest KK-odd-quark may be longer, given the analogous
spectrum, than that of a corresponding squark LSsP.

\subsection{Final Remarks}

Let us close with some remarks on the general search strategy that
might be adopted by the collider experiments seeking to find or
exclude these phenomena.  First, from the present work and
\cite{hidval, higgs}, it should be clear that there is more than ample
reason to make high-efficiency detection of highly displaced vertices an
urgent priority in the last year of preparation for the LHC.  Unlike
most processes that will be sought at the LHC, this is one with no
standard model background; all the backgrounds are instrumental.  As
such, it has rather different issues compared to those that trouble
many searches at the LHC.  In some scenarios,
highly displaced vertices may be the key phenomenon
allowing new signals to be identified and high-purity samples to be
obtained.  Comprehensive studies on a wide variety of highly-displaced
vertex signals do not yet seem to have been done, but would be
welcome.  Efforts to improve track-finding algorithms might be an
especially valuable use of resources.  

Another potentially important issue is the efficiency for detecting
large numbers of ordinary displaced vertices in events with many $b$
jets and perhaps $\tau$ leptons.  The classes of events considered
here may be very busy, with multiple hard jets along with the soft
jets from the v-hadron decays.  In this context, the challenge of
vertexing even with moderate efficiency seems daunting; but meeting
this challenge successfully may be essential for finding supersymmetry
or other new physics, and deserves additional consideration.

Given the vertexing capabilities of the LHCb detector, it would be
well worth exploring its sensitivity to these phenomena.  Despite the
fact that the LHCb detector cannot reconstruct entire events and
cannot therefore measure MET, as its coverage is limited to the
forward direction, it may be especially capable of detecting the LSsP
decays, or the ensuing v-hadron decays, some fraction of which will
occur within its acceptance.  Perhaps the LSsP and v-hadron decays
will first be seen and analyzed at LHCb; the information thereby
obtained will then be useful both for the CMS and ATLAS detectors in
their analysis of supersymmetric production events, and for general
efforts to understand the underlying supersymmetric model.

Finally, as should be evident, Tevatron limits on supersymmetric
particle production, and other similar models, need to be reconsidered
in this context.  A comprehensive search for events with
highly-displaced vertices might well reveal both a v-sector and the
first signs of supersymmetry, or perhaps something else entirely
unexpected.  The author hopes that this work, and that of
\cite{hidval, higgs}, will be sufficiently motivating for the Tevatron
experiments to undertake these difficult but important searches.

I thank A. Haas,  H. Frisch, H. Lubatti, A. Nelson, N. Weiner, and especially
K. Zurek for discussions.
This work was supported by U.S. Department
of Energy grant DE-FG02-96ER40956.

\end{document}